# Radio Resource Management Based on Reused Frequency Allocation for Dynamic Channel Borrowing Scheme in Wireless Networks

Mostafa Zaman Chowdhury[1], Mohammad Arif Hossain[2], Shakil Ahmed[1], and Yeong Min Jang[2]

[1]Dept. of Electrical and Electronic Engineering, Khulna University of Engineering & Technology, Khulna-9203, Bangladesh

[2]Dept. of Electronics Engineering, Kookmin University, Seoul 136-702, Korea

E-mail: mzceee@yahoo.com, dihan.kuet@gmail.com, shakileee076@gmail.com, yjang@kookmin.ac.kr

**Abstract—** In the modern era, cellular communication consumers are exponentially increasing as they find the system more user-friendly. Due to enormous users and their numerous demands, it has become a mandate to make the best use of the limited radio resources that assures the highest standard of Quality of Service (QoS). To reach the guaranteed level of QoS for the maximum number of users, maximum utilization of bandwidth is not only the key issue to be considered, rather some other factors like interference, call blocking probability etc. are also needed to keep under deliberation. The lower performances of these factors may retrograde the overall cellular networks performances. Keeping these difficulties under consideration, we propose an effective dynamic channel borrowing model that safeguards better QoS, other factors as well. The proposed scheme reduces the excessive overall call blocking probability and does interference mitigation without sacrificing bandwidth utilization. The proposed scheme is modeled in such a way that the cells are bifurcated after the channel borrowing process if the borrowed channels have the same type of frequency band (i.e. reused frequency). We also propose that the unoccupied interfering channels of adjacent cells can also be inactivated, instead of cell bifurcation for interference mitigation. The simulation endings show satisfactory performances in terms of overall call blocking probability and bandwidth utilization that are compared to the conventional scheme without channel borrowing. Furthermore, signal to interference plus noise ratio (SINR) level, capacity, and outage probability are compared to the conventional scheme without interference mitigation after channel borrowing that may attract the considerable concentration to the operators.

Keywords- Dynamic channel borrowing, Quality of Service (QoS), cell bifurcation, overall call blocking probability, bandwidth utilization, interference mitigation, SINR level, outage probability, and channel capacity.

## 1. Introduction

Wireless networks are probing for an optimum way so that it can keep pace with the fastest growing cellular communication consumers by ensuring maximum utilization of radio resource in spite of limited bandwidth spectrum, obviously retaining the highest standard of Quality of Service (QoS) at all the time [1]. In order to achieve the guaranteed QoS level, excessive call blocking probability, interference etc. may become hindrances.

Hence, due to interference problem, signal to interference plus noise ratio (SINR) level, and system capacity may decrease and may also increase in the outage probability that disturb a sound cellular communication [2-4].

In spite of congested traffic, there may be unoccupied channels in any cell of a cluster that depends upon the total call arrival rate of the network [5]. If a model can be developed that ensures maximum utilization of these unoccupied channels, this model can fascinate to fulfill the need of the excessive consumers in wireless communication. In cellular networks, if there is excessive traffic in a cell compared to the number of the original channels, the cell has the opportunity to borrow the required number of channels from adjacent cells if the cells have unoccupied channels. In this paper, we propose a dynamic channel borrowing scheme in an effectual way. We consider three types of reused frequency band in a cluster of seven cells. An algorithm is proposed for selection of adjacent cells having maximum number of available channels and the process for channel borrowing. In our model, we define the cell as reference cell where the traffic intensity is the highest among the cells in the cluster. The required numbers of channels of the reference cell are borrowed from the adjacent cell or cells of the cluster where the maximum numbers of unused channels are available.

We propose that for interference management, the cells are bifurcated, named inner part and outer part [6-7] (according to necessity, in next section, we discuss it broadly). The novelty of our proposed model is the proposal of various techniques for interference mitigation after dynamic channel borrowing process. The borrowed channels are provided to the inner part users of the reference cell. Moreover, if the channels of same frequency band in the adjacent cells cause interference with the channels of the reference cell, then the interfering channels of the adjacent cells are also provided to the inner part users of those adjacent cells to reduce interference.

In the previous time, fixed channel assignment architecture (FCA) [8] and hybrid channel assignment architecture (HCA) [9] have been proposed that show better performance in case of bandwidth utilization without considering interference management. A channel borrowing scheme without locking (CBWL) model [10] has been proposed which shows that the cells of the system borrow channels from its adjacent cells when calls arrive but these channels cannot be served by the normal channels. If the channels are borrowed from the other cells, the cells use it with such reduced power so that the borrowed channels are not necessary to be locked. Dynamic channel allocation with interference mitigation architecture [11] and interference declination approach for OFDMA networks [12-13] describe some interference management techniques. Our proposed model shows better performances compared to the above models.

In the proposed architecture, the channel borrowing procedure is done dynamically. It means either the available channels (i.e. if the available channels are less than the required number of channels) or the required numbers of channels (i.e. if the available channels are more than the required channels) are borrowed from the adjacent cells by the reference cell. However, for interference management, cell bifurcation is done for both reference cell and adjacent cells according to condition of the cells (we broadly discuss in our proposed system model). Furthermore, the interfering channels of the adjacent cells are also inactivated to minimize interference.

In case of interference management, three types of scheme are proposed in our paper. Firstly, interference management is done by the bifurcation of the reference cell. The frequency band that causes interference is provided to the inner part users. Secondly, both the reference cell and the adjacent cells are bifurcated. Finally, instead of reference cell and adjacent cells bifurcation, the unused channels with same frequency band that may cause interference, is made inactive in the adjacent cells, thus interference mitigation is made possible. These approaches result in interference declination, higher system capacity, and lower outage probability for a cellular network where there exists excessive traffic. The proposed model also shows convenient performance that reduces excessive overall call blocking probability and ensures maximum bandwidth utilization which may draw the attention of the operators of the cellular networks.

The rest of this paper is organized as follows: Section 2 shows the proposed dynamic channel borrowing scheme with algorithm and the proposed interference declination scheme with proper illustration. Call blocking probability using the queuing analysis for the proposed scheme is shown in Section 3. In Section 4, the capacity and outage probability analysis of the proposed scheme are shown. The performance evaluation of our proposed model is demonstrated in Section 5. Finally, summary and concluding notes are drawn in Section 6.

## 2. Proposed Dynamic Channel Borrowing and Interference Mitigation Scheme

Existing and upcoming wireless networks are requisite to support the maximum number of users within limited resources. Due to excessive rise of the users in wireless networks, the proposed dynamic channel borrowing model can be a useful way to compensate the demand. We propose a dynamic channel borrowing scheme as well as interference mitigation process through which maximum bandwidth utilization with less overall call blocking probability and better QoS are ensured. In this system model, we use the basic nomenclatures that are defined in Table 1.

**Table 1.** Basic nomenclature for System model

| Symbol | Definition |
|---|---|
| $X$ | The frequency band allocated for reference cell |
| $Y$ | The frequency band allocated for each of cell 2, cell 4, and cell 6 |
| $Z$ | The frequency band allocated for each of cell 3, cell 5, and cell 7 |
| $Y'$ | The frequency band which can be borrowed by the reference cell from cell 2 or cell 4 or cell 6 |
| $Y'_4$ | Fraction of the frequency band $Y'$ that is provided to the inner part users of cell 4 |
| $Y'_6$ | Fraction of the frequency band $Y'$ that is provided to the inner part users of cell 6 |
| $Z'$ | The frequency band which can be borrowed by the reference cell from cell 3 or cell 5 or cell 7 |
| $Z'_5$ | Fraction of the frequency band $Z'$ that is provided to the inner part users of cell 5 |
| $Z'_7$ | Fraction of the frequency band $Z'$ that is provided to the inner part users of cell 7 |
| $Y_{Th}$ | The maximum frequency band which can be borrowed from $Y$ frequency band |
| $Z_{Th}$ | The maximum frequency band which can be borrowed from $Z$ frequency band |
| $N$ | Total number of original channels in each cell |
| $N_{req}$ | Total number of channels that are required for meeting the demand of excessive users in the reference cell |
| $N_{av,m}$ | Total number of unused channels in cell $m$, where $1 \leq m \leq 7$ |
| $A_{max}$ | The cell having maximum number of unoccupied channels among group $A$ (i.e. cell 2, cell 4, and cell 6) |
| $B_{max}$ | The cell having maximum number of unoccupied channels among group $B$ (i.e. cell 3, cell 5, and cell 7) |
| $N_{req,add}$ | The additional number of channels which are required to borrow after borrowing $N_{av,m}$ |

**2.1. System Model**

In our proposed scheme, we consider a cluster of seven cells where three types of frequency band are reused, named $X$, $Y$, and $Z$. We assume $|X|=|Y|=|Z|$. Frequency allocation of these cells before dynamic channel borrowing process is shown in Fig. 1. Each cell encompasses $N$ number of channels before the borrowing process. We consider two groups of cells in the cluster based on the reused frequency allocation for channel borrowing process, named group $A$ and group $B$. Group $A$ consists of cell 2, cell 4, and cell 6 that has same frequency band. Moreover, group $B$ contains cell 3, cell 5, and cell 7 which has also same type of frequency band. We define the cells as $A_{max}$ and $B_{max}$ where the number of unoccupied channels are maximum in group $A$ and group $B$, respectively. Besides, we consider the number of unoccupied channels in the cell $A_{max}$ is greater than the number of unoccupied channels in the cell $B_{max}$. Figure 1 illustrates that cell 1 has channels with frequency band $X$, same type of frequency band named $Y$ is contained by cell 2, cell 4, and cell 6 while cell 3, cell 5, and cell 7 cover frequency band $Z$ individually. Bandwidth spectrum of the cells corresponding to Fig. 1 is represented in Fig. 2.

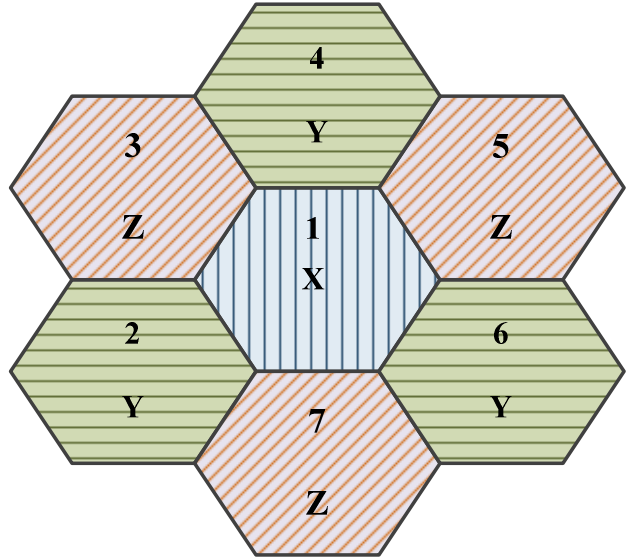

**Fig. 1.** Frequency band allocation in the cells of a cluster before dynamic channel borrowing process.

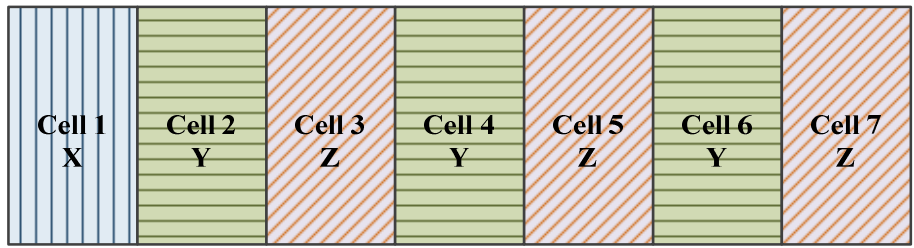

**Fig. 2.** Frequency band of channels in the cells before dynamic channel borrowing process.

According to our system model, cell 1 is the reference cell as it is assumed that the traffic intensity in the cell is higher than the traffic intensity of other six cells of the cluster. Consequently, there emerges an inevitability to borrow channels from the adjacent cells. The system model clarifies that in the first search for unoccupied channels, if the cell $A_{max}$ (say cell 2 belongs to group $A$) is found with the maximum number of unoccupied channels, the reference cell borrows channels unceasingly from cell 2 till the channels are required for the reference cell or borrows the available channels. Subsequently, if the reference cell is in need of more channels, then it borrows channels from the cell $B_{max}$ of other group (say cell 3 belongs to group $B$) which has the maximum number of unoccupied channels at that instant. This is simply demonstrated in Fig. 3. It shows that the frequency band $Y'$ and $Z'$ are borrowed by the reference cell from cell 2 and cell 3, respectively. Thus, the frequency band of cell 2 and cell 3 are reduced to ($Y-Y'$) and ($Z-Z'$), respectively. This results in the escalation of the frequency band from $X$ to $(X+Y'+Z')$ of the cell 1. As a result, the users of the reference cell face interference from cell 4, cell 5, cell 6, and cell 7 as no interference management is done. Figure 4 reveals the reallocated frequency band corresponding to the dynamic channel borrowing process of Fig. 3. As the frequency bands ($X$, $Y$, and $Z$) of the cells in the cluster are reused, there creates interference problem due to the channels of the adjacent cells to the users of the reference cell.

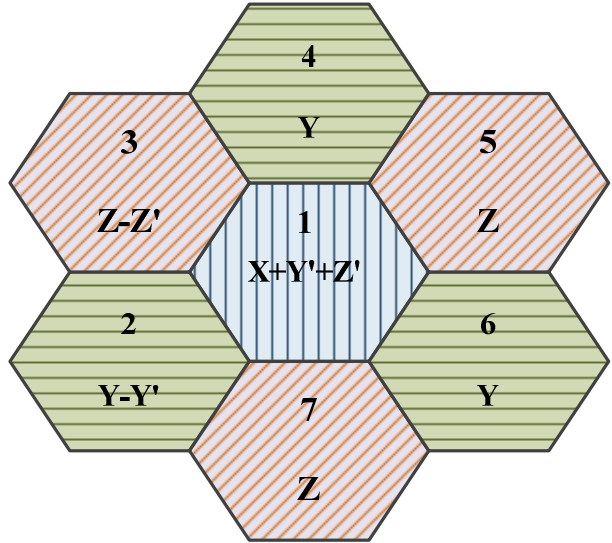

**Fig. 3.** Frequency band allocation in the cells after dynamic channel borrowing process without interference mitigation.

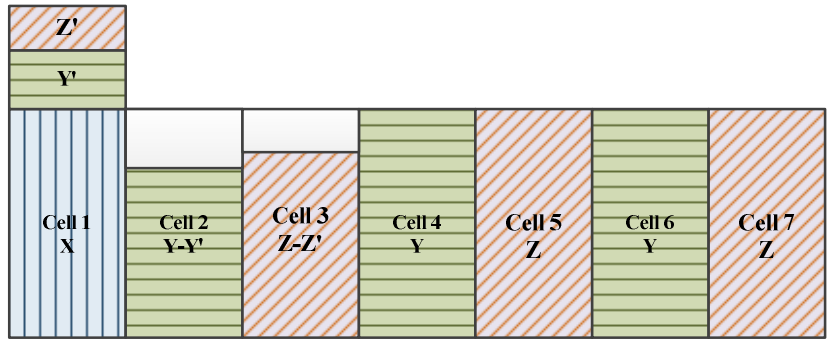

**Fig 4.** Reallocated frequency band of the cells after dynamic channel borrowing process.

**2.2. Borrowing Architecture**

When the original channels of reference cell become occupied by the existing users, the cell needs more channels to keep the newly arrived traffic active. The proposed scheme shows the borrowing process in which the channels can be borrowed from the adjacent cell of one group (we consider group $A$) at first. If the reference cell needs more channels, it can borrow channels from cell of group $B$. According to our proposed model, $N_{req}$ is total number of channels that are required for meeting the demand of excessive consumers in the reference cell and $N_{av,m}$ is the total number of unoccupied channels in the cell $m$, where $2 \leq m \leq 7$. The dynamic channel borrowing process involves an exploration for $A_{max}$ in cell 2, cell 4, and cell 6 while search for $B_{max}$ in cell 3, cell 5, and cell 7. In the system model, we assume cell 2 as $A_{max}$ for group $A$ and cell 3 as $B_{max}$ for group $B$.

The reference cell (cell 1) borrows $N_{req}$ channels from the cell $A_{max}$ (cell 2). If $N_{req}$ channels are less than $N_{av,2}$ (the number of unoccupied channels available in cell 2) channels then cell 1 borrows $N_{req}$ channels and the total number of channels of cell 1 becomes $N+N_{req}$. However, if $N_{req}$ is greater than $N_{av,2}$, then cell 1 borrows the total number of unoccupied channels $N_{av,2}$. Hence, the total number of channels of cell 1 becomes $N+N_{av,2}$. If the total required number of channels $N_{req}$ is not achieved, the reference cell can borrow the required number of channels from the cell $B_{max}$ (i.e. cell 3). Let, $N_{req,add}$ is the additional number of required channels to obtain $N_{req}$ channels after borrowing from cell 2. So, we can write $N_{req,add} = N_{req}-N_{av,2}$. If $N_{req,add}$ is less than $N_{av,3}$ (the number of unused channels available in cell 3), then cell 1 borrows $N_{req,add}$ channels and the total number of channels of cell 1 becomes $N+N_{av,2}+N_{req,add}$. If $N_{req,add}$ is greater than $N_{av,3}$, then cell 1 borrows $N_{av,3}$ channels and the total number of channels of the reference cell becomes $N+N_{av,2}+N_{av,3}$.

The mechanism for the selection of the adjacent cells and the borrowing process is proposed in algorithm I. The terms used in the proposed algorithm are defined in Table 1.

**Algorithm I:** Mechanism for selection of adjacent cells and channel borrowing process

1: **while** a call arrives at reference cells (cell 1) **do**
2:   **if** $N_{req}>N_{av,1}$ **then**
3: select a cell from group $A$ to borrow channels
4:   **if** $N_{av,m} \geq N_{req}$ **then**
5:   **if** $N_{av,2}>N_{av,4}$ **then**
6:   **if** $N_{av,2}>N_{av,6}$ **then**
7:       select cell 2 as $A_{max}$
8:   **else**
9:       select cell 6 as $A_{max}$
10:   **end if**
11:   **else**
12:   **if** $N_{av,4}>N_{av,6}$ **then**
13:       select cell 4 as $A_{max}$
14:   **else**
15:       select cell 6 as $A_{max}$
16:   **end if**
17:   **end if**
18: select cell $A_{max}$ (i.e. cell 2) to borrow channels
19:   **if** $N_{req} \leq N_{av,2}$ **then**
20:       borrow $N_{req}$ channels
21:   **else**
22:       borrow $N_{av,2}$ channels from $A_{m,uc}$ and additional required number of channels from the cell $B_{max}$ of group $B$ if available
23:   **if** $N_{av,m} \geq (N_{req}-N_{av,2})$ **then**
24:   **if** $N_{av,3}>N_{av,5}$ **then**
25:   **if** $N_{av,3}>N_{av,7}$ **then**
26:       select cell 3 as $B_{max}$
27:   **else**

28:         select cell 7 as $B_{max}$
29:     **end if**
30:   **else**
31:     **if** $N_{av,5} > N_{av,7}$ **then**
32:         select cell 5 as $B_{max}$
33:     **else**
34:         select cell 7 as $B_{max}$
35:     **end if**
36:   **end if**
37: select cell $B_{max}$ (i.e. cell 3) and borrow ($N_{req}$-$N_{av,2}$) channels
38:   **if** ($N_{req}$-$N_{av,2}$)≤$N_{av,3}$  **then**
39:       borrow ($N_{req}$-$N_{av,2}$) channels
40:   **else**
41:       borrow $N_{av,3}$ channels
42:   **end if**
43:   **else**
44:       borrow no channel from $B_{max}$ cell
45:   **end if**
46:  **end if**
47:  **else**
48:      borrow no channel from $A_{max}$ cell
49:  **end if**
50:  **else**
51:      borrow no channel
52: **end if**
53: **end while**

### 2.3. Interference Management

Frequency reusing techniques in the cellular networks cause the interference problem which results in lower system capacity and lower SINR level as well as higher outage probability [14]. To mitigate interference problems, we offer three unlike process which can be very useful to enhance QoS of the cellular wireless networks.

*Bifurcation of reference cell*

When the reference cell borrows channels $Y'$ and $Z'$ from cell 2 and cell 3, respectively, the users of the cell 1 receive interference due to $Y'$ channels of cell 4 and cell 6. Similar occurrence can be happened to the users due to $Z'$ channels of cell 5 and cell 7. The process of interference mitigation involves the bifurcation of the reference cell as shown in Fig. 5(a). According to the proposed scheme, the borrowed frequency bands ($Y'$ and $Z'$) from cell 2 and cell 3 are allocated to the inner part users and the outer part users are provided with the frequency band $X$ of cell 1. Thus the approach solves the interference problem to a great extent. In the meantime, the bandwidth of the reference cell increases and the increased bandwidth can accommodate the excessive traffic.

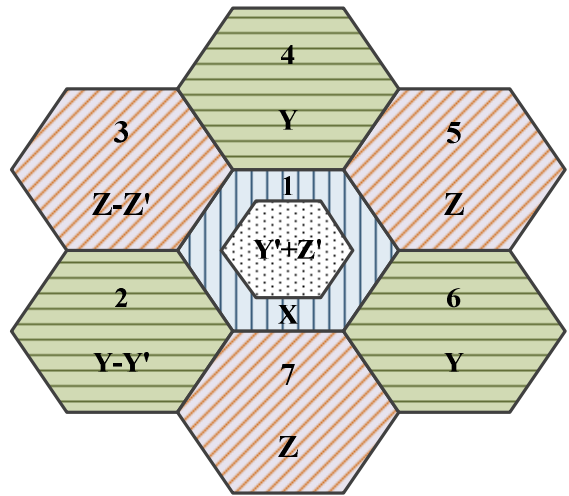

(a)

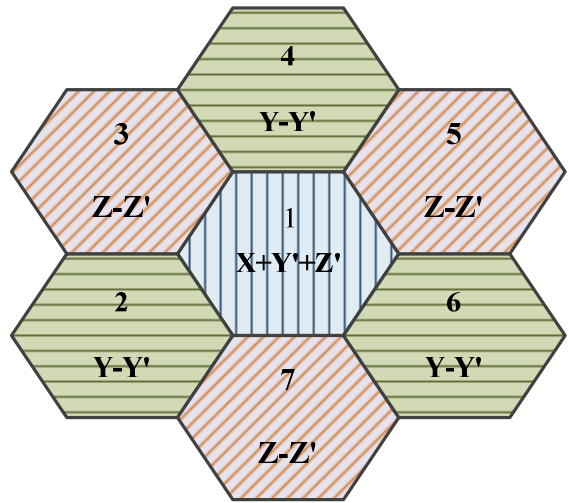

(b)

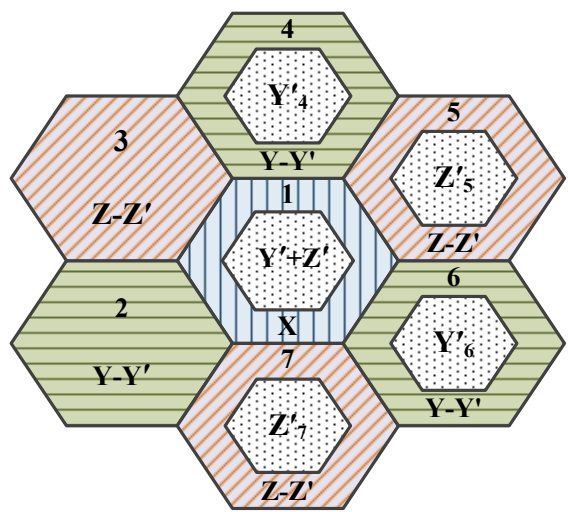

(c)

**Fig. 5.** Resource allocation after dynamic channel borrowing process with interference mitigation techniques (a) bifurcation process of reference cell (b) blockage of the interfering channels of adjacent cells (c) bifurcation process of reference cell and adjacent cells.

*Keeping interfering channels blocked*

Another approach of the proposed scheme points towards the process of making the interfering channels of the adjacent cells (from which channels are not borrowed) inactive when the reference cell borrows channels from cell 2 and cell 3. This procedure is possible when the interfering channels of the adjacent cells remain unused. Figure 5(b) illustrates the interference mitigation technique. If $Y'$ channels of cell 4 and cell 6 as well as $Z'$ channels of cell 5 and cell 7 remain unused, the interfering channels can be blocked. Consequently, interference can be reduced to a great extent.

*Bifurcation of adjacent cells*

The last tactic for interference mitigation has been proposed which includes the bifurcation of the adjacent cells that lessens the sophistication of the system with creditable performance. This approach can be applied if all or partial part of the interfering channels is occupied in the adjacent cells. The bifurcation process has been illustrated in Fig. 5(c). In this manner, if $Y'$ and $Z'$ channels are borrowed from cell 2 and cell 3, respectively by the reference cell, then the reference cell and other adjacent cells are bifurcated and the interfering channels are organized in such a way that interference can be reduced to a considerable level. The borrowed $Y'$ and $Z'$ channels are allocated to the inner part users of the reference cell.

We assume that $Y'_4$ and $Y'_6$ frequency bands are occupied in cell 4 and cell 6, respectively. Here, $Y'_4$ and $Y'_6$ indicate the total or partial part of $Y'$. The ranges of $Y'_4$ and $Y'_6$ lie between 0 and $Y'$ (i.e. $0 \leq Y'_4 \leq Y'$ and $0 \leq Y'_6 \leq Y'$). The occupied $Y'_4$ frequency band of cell 4 and $Y'_6$ frequency band of cell 6 may be equal or not. To reduce the interference of the reference cell, $Y'_4$ and $Y'_6$ frequency bands are provided to the inner part users of cell 4 and cell 6, respectively. Similarly, $Z'_5$ and $Z'_7$ frequency bands are occupied in cell 5 and cell 7, respectively where, $Z'_5$ and $Z'_7$ frequency bands indicate the total or partial part of $Z'$. 0 to $Z'$ is the limit for both frequency band of $Z'_5$ of cell 5 and $Z'_7$ of cell 7 (i.e. $0 \leq Z'_5 \leq Z'$ and $0 \leq Z'_7 \leq Z'$). The occupied $Z'_5$ frequency band of cell 5 and $Z'_7$ frequency band of cell 7 may be equal or not. Now, $Z'_5$ and $Z'_7$ frequency bands are provided to the inner part users of cell 5 and cell 7, respectively. So, it is easy to comprehend that if the unoccupied frequency band among $Y'$ of cell 4 and cell 6 as well as $Z'$ of cell 5 and cell 7 become occupied, those frequency bands are delivered to the inner part users of the respective cells. According to the process, $Y'_4$ and $Y'_6$ frequency bands of cell 4 and cell 6, respectively can be extended up to $Y'$ frequency band according to the requirement of the users. Similarly, $Z'_5$ and $Z'_7$ frequency bands of cell 5 and cell 7, respectively can be extended up to $Z'$ frequency band according to the necessity of the inner part users. However, if cell 2 and cell 3 need more channels to keep the newly arrived users active after borrowing channels by the reference cell, then cell 2 and cell 3 can reuse the total or partial part of the frequency bands $Y'$ and $Z'$, respectively. In this case, to avoid interference the frequency band $Y'$ and $Z'$ are provided to the inner part users of cell 2 and cell 3, respectively.

When the channels are borrowed from cell 2, then the channels of cell 4 and cell 6 ranges to the same frequency band of the borrowed channels cause interference if they get occupied. Consequently, these occupied interfering channels are provided to the inner part consumers of cell 4 and cell 6. Likewise, for cell 3, the interfering cells are cell 5 and cell 7 and same actions may be implemented in the cells of group *B* according to our proposed scheme.

If channels are available in the adjacent cells of same group (i.e. either group *A* or group *B*), then the channels can be borrowed up to threshold value $Y_{Th}$ or $Z_{Th}$) from such cell which has the maximum number of unoccupied channels. Besides, if more channels are required, then the reference cell can borrow channels from the adjacent cells of group *B*. However, only the occupied channels of the adjacent cells which may be the total or partial part of the channels that range to same frequency band of the borrowed channels are provided to the inner part users.

## 3. Queuing Analysis

The proposed scheme can be modeled as *M/M/K/K* queuing system. The call arriving process is assumed to be Poisson. It is assumed that the arrival process of the traffic in the cellular networks requires exponentially distributed amount of service [15]. The ending of a call results in the departure of the call from the system. Table 2 shows the basic nomenclatures for the analysis of the paper.

**Table 2.** Nomenclature for queuing analysis, capacity and outage probability analysis

| Symbol | Definition |
|---|---|
| $M$ | Total number of cells in a cluster |
| $m$ | Cell *m* in the cluster |
| $\lambda_m$ | Call arrival rate of cell *m* |
| $1/\mu_m$ | Channel holding time of cell *m* |
| $N_m$ | Total number of channels in cell *m* before channel borrowing process |
| $N'_m$ | Total number of channels in cell *m* after channel borrowing process |
| $N_{Th}$ | Threshold value for borrowing channels |
| $P_m(i)$ | Steady state probability of state *i* in cell *m* |
| $P_{Bm}$ | Call blocking probability in cell *m* |
| $P_{B_T}$ | Overall call blocking probability of the system |
| $f_c$ | Center frequency |
| $h_b$ | Height of the base station (BS) |
| $h_m$ | Height of the mobile antenna (user) |
| $d$ | Distance between the transmitter and the receiver |
| $S_o$ | Received signal power from BS |
| $P_{out}$ | Outage probability of the system |
| $R$ | Maximum number of interfering tiers in the network |
| $r$ | Any interfering tier in the network |
| $N_r$ | Maximum number of interfering cells in *r*-th tier |
| $i_r$ | An interfering cell in the *r*-th tier |
| $I_r(i_r)$ | Received power of interference from *i*-th cell of *r*-th tier |
| $L$ | Path loss exponent for cellular networks |
| $L_{OW}$ | Penetration loss |
| $C$ | Capacity of the system |

Figure 6 designates the Markov chain before dynamic channel borrowing process while Fig. 7 stands for the Markov chain of the proposed scheme. Suppose, $M$ is the maximum number of cells and $m$ refers cell $m$ in the cluster of the system. Maximum number of calls that can be accommodated in cell $m$ is $N_m$. $N_{Th}$ represents the threshold value below which channels cannot be borrowed. In the system, $\lambda_m$ and $\mu_m$ represent the call arrival rate and channel release rate for cell $m$, respectively. We assume that each cell contains $N$ number of channels before channel borrowing. We can write $N_1 = N_2 = N_3 = ... = N_m = ... = N_M = N$

However, $N_m$ and $N'_m$ represent the total number of channels in cell $m$ before and after channel borrowing process, respectively. According to our system model, the reference cell can borrow channels from cell $m$ up to the threshold value. For reference cell, $N'_m$ is greater than $N_m$ while for the adjacent cells from which the channels are borrowed, $N'_m$ is less than or equal to $N_m$. $N'_m$ is equal to $N_m$ for the cells from which channels are not borrowed i.e. for reference cell $N'_1 > N_1$ and for adjacent cells $N_{Th} \leq N'_m \leq N_m$ when $2 \leq m \leq 7$.

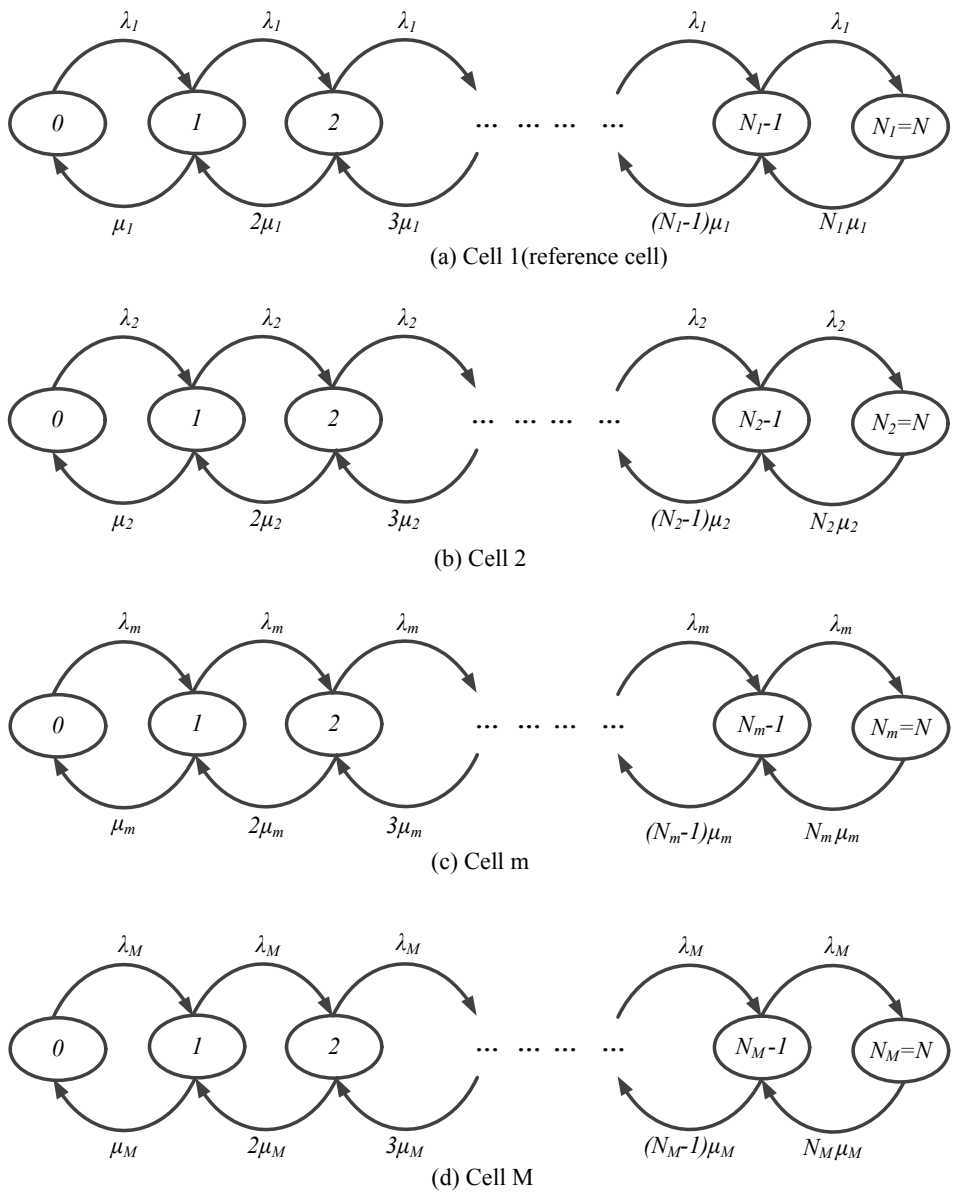

**Fig. 6.** Markov chain before channel borrowing process.

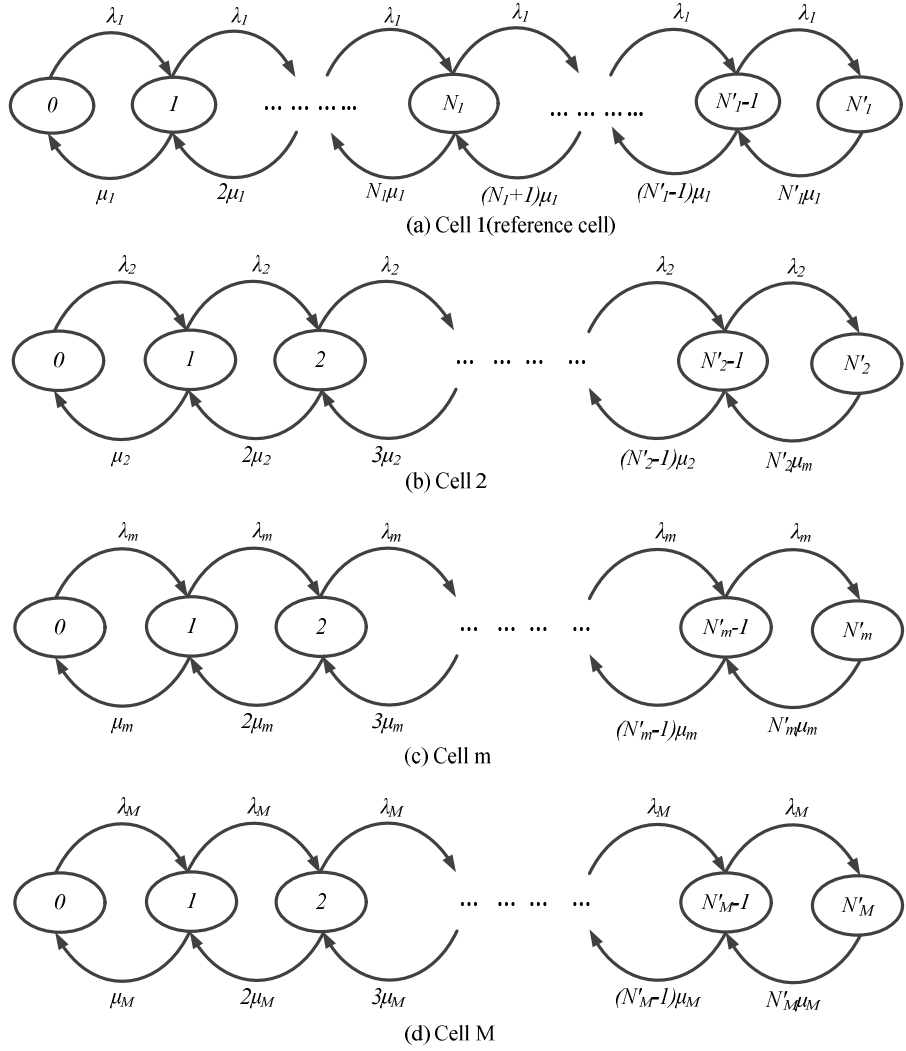

**Fig. 7.** Markov chain for the proposed model.

Let, $P_m(i)$ is the steady state probability of state $i$ of cell $m$ after channel borrowing process. Thus, the accumulated probability for each state of cell $m$ can be expressed as

$$\sum_{i=0}^{N'_m} P_m(i) = 1 \qquad (1)$$

The steady state probability of state $i$ of cell $m$ can be stated as

$$P_m(i) = \frac{(\lambda_m)^i}{i!(\mu_m)^i} P_m(0), \quad 0 \leq i \leq N'_m \qquad (2)$$

From normalizing conditions, we have

$$P_m(0) = \left[\sum_{i=0}^{N'_m} \frac{(\lambda_m)^i}{i!(\mu_m)^i}\right]^{-1} \qquad (3)$$

The call blocking probability of cell $m$ is $P_{Bm}$ and the overall call blocking probability of the system is $P_{BT}$.

From (1) to (3), $P_{Bm}$ and $P_{BT}$ can be calculated as (4) and (5), respectively, shown below.

$$P_{Bm} = \frac{(\lambda_m)^{N'_m}}{N'_m!(\mu_m)^{N'_m}} P_m(0) \tag{4}$$

$$P_{BT} = 1 - \frac{\sum_{m=1}^{M} \lambda_m(1 - P_{Bm})}{\sum_{m=1}^{M} N'_m} \tag{5}$$

The bandwidth utilization of the system can be expressed as

$$B_W = \frac{(1 - P_{BT})(\sum_{m=1}^{M} \lambda_m)}{\mu_T \sum_{m=1}^{M} N'_m} \tag{6}$$

where $\mu_T$ denotes the average channel release rate of the system.

## 4. SINR, Capacity, and Outage Probability Analysis

For interference management, there are many strategies available at present in the cellular networks. Mitigation of interference and noise is one of the biggest challenges because these strictly decrease the performances in terms of outage probability, SINR level, and also the capacity of the wireless link [16]. Calculation of signal coverage for base signal and interference are obligatory for sound implementation and efficient output of the cellular system. The signal coverage is greatly influenced by the radio frequency and topography. Okumura-Hata model have been used for cellular path loss calculation [17]. From the model, we can write,

$$L = 69.55 + 26.16 \log f_c - 13.82 \log h_b - a(h_m) + (44.9 - 6.55 \log h_b) \log d + L_{OW} \quad \text{[dB]} \tag{7}$$

$$a(h_m) = 1.1(\log f_c - 0.7)h_m - (1.56 \log f_c - 0.8) \quad \text{[dB]} \tag{8}$$

where $L$ is the path loss exponent for cellular networks, $f_c$ is the center frequency of the macrocell, $h_m$ depicts the height of mobile antenna, $h_b$ is the height of BS, $L_{OW}$ is the penetration loss, and $d$ is defined as the distance between the BS and the MS.

Considering the spectrum of transmitted signals is spreads, we can approximate the interference as AWGN [18]. In case of capacity, we can write Shannon capacity formula that can be used for ease of calculation as

$$C = \log_2(1 + SINR) \quad \text{[bps/Hz]} \tag{9}$$

The capacity of a wireless communication decreases with reduced SINR level. Figure 8 interprets the various signals and interference that immensely affect the capacity and the outage probability of a cellular network.

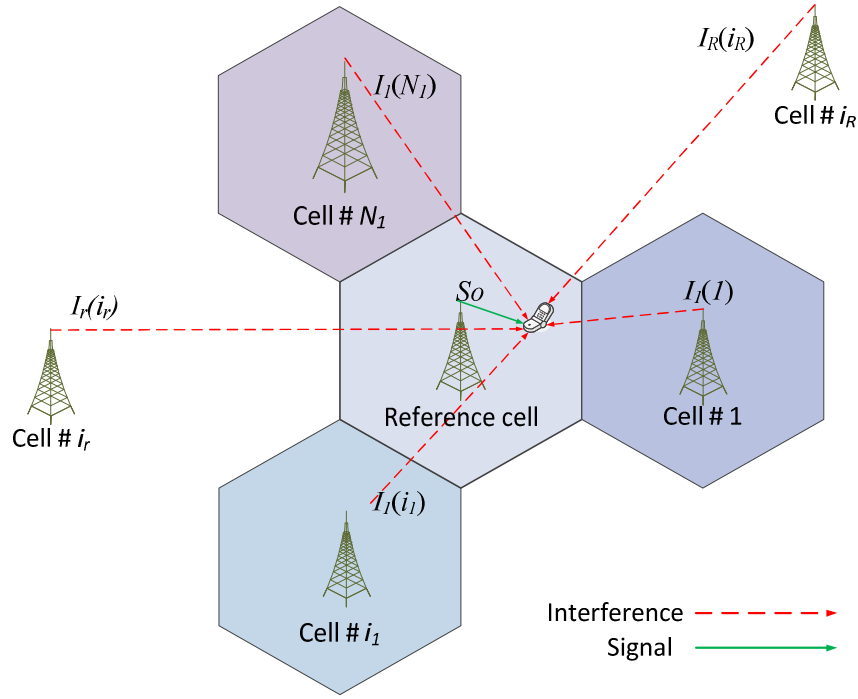

**Fig. 8**. Scenario of signals and interferences for a consumer situated in the reference cell of a cellular network.

The received SINR level for users can be expressed as

$$SINR = \frac{S_o}{\sum_{i_1=1}^{N_1} I_1(i_1) + \sum_{i_2=1}^{N_2} I_2(i_2) + ... + \sum_{i_r=1}^{N_r} I_r(i_r) + ... + \sum_{i_R=1}^{N_R} I_R(i_R)}$$

$$= \frac{S_o}{\sum_{T=1}^{R}\left\{\sum_{i_T=1}^{N_T} I_T(i_T)\right\}} \quad (10)$$

where $S_0$ is the received power signal from the BS, $r$ represents a tier among the interfering tiers. $i_r$ and $N_r$ refer a cell among the interfering cells of $r$-th tier and maximum number of interfering cells in $r$-th tier, respectively. $I_r(i_r)$ is the received power of the interference from $i_r$-th cell of $r$-th tier. For reduced outage probability, we can write the formula [19] as

$$P_{out} = P_r(SINR < \gamma) \quad (11)$$

where $\gamma$ is the threshold value of SINR level. We consider the value of $\gamma$ in our performance analysis as 9 dB. If its value goes below this level, then there is no acceptable reception. Consecutively, we can write from (11)

$$P_{out} = P_r\left(\frac{S_o}{\sum_{T=1}^{R}\left\{\sum_{i_T=1}^{N_T} I_T(i_T)\right\}} < \gamma\right)$$

$$= P_r\left\{\sum_{T=1}^{R}\left\{\sum_{i_T=1}^{N_T} I_T(i_T)\right\} > \left(\frac{S_o}{\gamma}\right)\right\} \quad (12)$$

For the proposed model, considering the interfering channels of the cells, the outage probability can be expressed as

$$P_{out} = 1 - \exp\left[-\frac{\gamma}{S_o}\sum_{T=1}^{R}\left\{\sum_{i_T=1}^{N_T}I_T(i_T)\right\}\right]$$

$$= 1 - \exp\left[-\frac{\gamma}{S_o}\left\{\sum_{i_1=1}^{N_1}I_1(i_1) + \sum_{i_2=1}^{N_2}I_2(i_2) + ... + \sum_{i_r=1}^{N_r}I_r(i_r) + ... + \sum_{i_R=1}^{N_R}I_R(i_R)\right\}\right]$$

$$= 1 - \left\{\prod_{i_1=1}^{N_1}e^{\left\{-\frac{\gamma}{S_o}I_1(i_1)\right\}Z(i_1)}\right\}\left\{\prod_{i_2=1}^{N_2}e^{\left\{-\frac{\gamma}{S_o}I_2(i_2)\right\}Z(i_2)}\right\}...\left\{\prod_{i_r=1}^{N_r}e^{\left\{-\frac{\gamma}{S_o}I_r(i_r)\right\}Z(i_r)}\right\}...\left\{\prod_{i_R=1}^{N_R}e^{\left\{-\frac{\gamma}{S_o}I_R(i_R)\right\}Z(i_R)}\right\}$$

$$= 1 - \prod_{T=1}^{R}\left\{\prod_{i_T=1}^{N_T}e^{\left\{-\frac{\gamma}{S_o}I_T(i_T)\right\}Z(i_T)}\right\}$$

(13)

Equation (13) can be used for the calculation of the outage probability of the system. $Z(i_T)$ indicates a binary function. The value of $Z(i_T)$ is associated with the allocated frequency band for the reference cell and the $i_T$-th cell of $T$-th tier. $Z(i_T)$ becomes 1 when the reference cell and the adjacent cells use same frequency band after borrowing channels. Otherwise, $Z(i_T)$ becomes zero. Insufficient interference management reduces the consumers QoS with aggrandizement of outage probability where appropriate interference mitigation solves the limitations. The reduced outage probability is expected for an efficient wireless network communication [20]. The bifurcation of cell reduces the $P_{out}$ and improves the SINR level and the capacity of the system.

## 5. Performance Evaluation

In this section, we analyze the performances of the proposed scheme. Hence, we study the outcome, considering the call arrival process to be Poisson. Overall call blocking probability and overall bandwidth utilization of the system depend on the individual call blocking probability and bandwidth utilization of the cells of a cluster. We illustrate the performances in terms of overall call blocking probability, overall bandwidth utilization, SINR level, capacity, and the outage probability. Overall bandwidth utilization and overall call blocking probability of the proposed model are compared to such model where dynamic channel borrowing process is absent. Besides, SINR level, outage probability, and the system capacity of the proposed scheme are compared to the approach where dynamic channel borrowing process is present but interference management is neglected. We assume the empirical ratio of call arrival rate of the seven cells in the cluster as 7:1:2:4:5:5:6. Total call arrival rate is defined as the summation of the respective call arrival rates of the cells in the cluster. Moreover, we consider only 1st and 2nd tier for the calculation of interference. Table 3 inclines the elementary parameters that are used for the performance analysis of the proposed scheme.

**Table 3:** Summary of the parameter values used in the analysis

| Parameter | Value |
|---|---|
| Number of cells in each cluster | 7 |
| Number of original channels in each cell | 100 |
| Number of reused frequency band | 3 |
| Threshold value of channels for borrowing in each cell | 70 |
| Center frequency | 1800 MHz |
| Transmitted signal power by the BS | 1.50 kw |
| Height of the BS | 100 m |
| Average channel holding time | 90 sec |
| Cell radius | 1 km |
| Penetration loss | 20 dB |
| Threshold value of SINR ($\gamma$) | 9 dB |
| Height of the mobile antenna | 5 m |

Figure 9 associates the overall call blocking probability between conventional model and the proposed model. The figure demonstrates the superiority of the proposed scheme as it reduces the overall call blocking probability of the system. At lower total call arrival rate, the traffic intensity in each cell is less compared to the original number of channels. So, all the cells in the cluster have sufficient channels to manage the traffic. Thus the overall call blocking probability of the proposed scheme shows approximate equal performance. As the total call arrival rate increases, the reference cell has to provide the excessive number of traffic by dynamically borrowing channels from adjacent cells. Consequently, the overall call blocking probability of the proposed model reduces meaningfully as the extra number of traffic is not blocked due to utilization of the unused channels of the adjacent cells. Thus the attractive performance proves the feasibility of our proposed scheme that may draw the attention of the operators for future wireless networks.

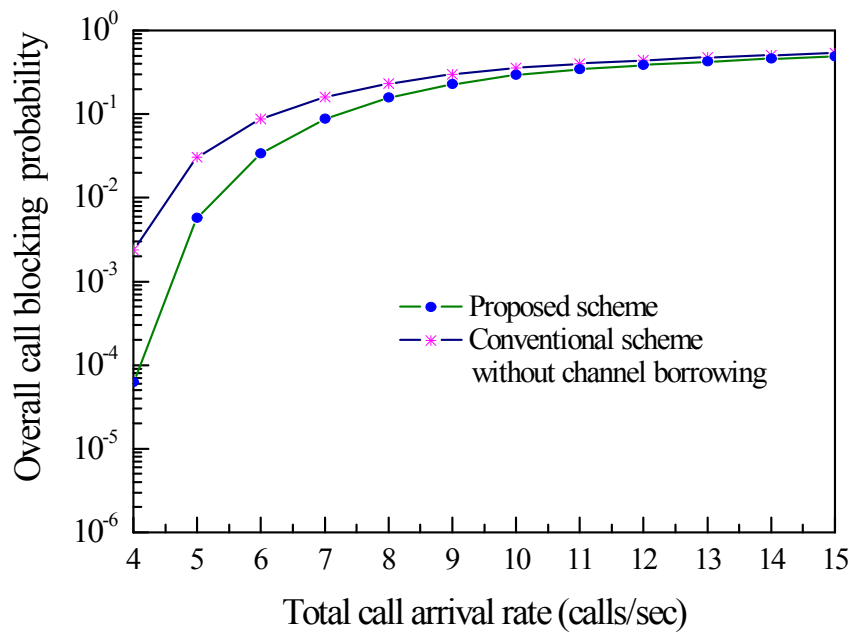

**Fig. 9.** Comparison of overall call blocking probability.

Figure 10 shows the comparison of the call blocking probability of the reference cell for the proposed scheme and the conventional model i.e. model without channel borrowing process. As the dynamic channel borrowing scheme confirms the activation of the excessive traffic and the conventional method fails to activate, verily the call blocking probability of the reference cell of the proposed model shows greatly better performance compare to the conventional model. The call blocking probability of a cell is higher if the call arrival rate of the cell is higher. In spite of higher call arrival rate in the reference cell, ensuring the lower call blocking probability proves nothing but the novelty of the proposed scheme.

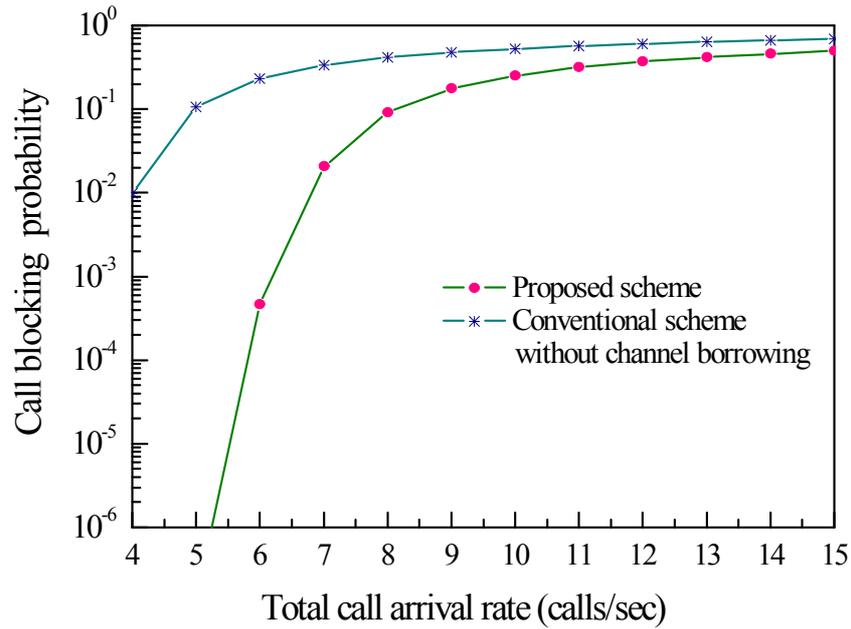

**Fig. 10.** Comparison of call blocking probability of the reference cell.

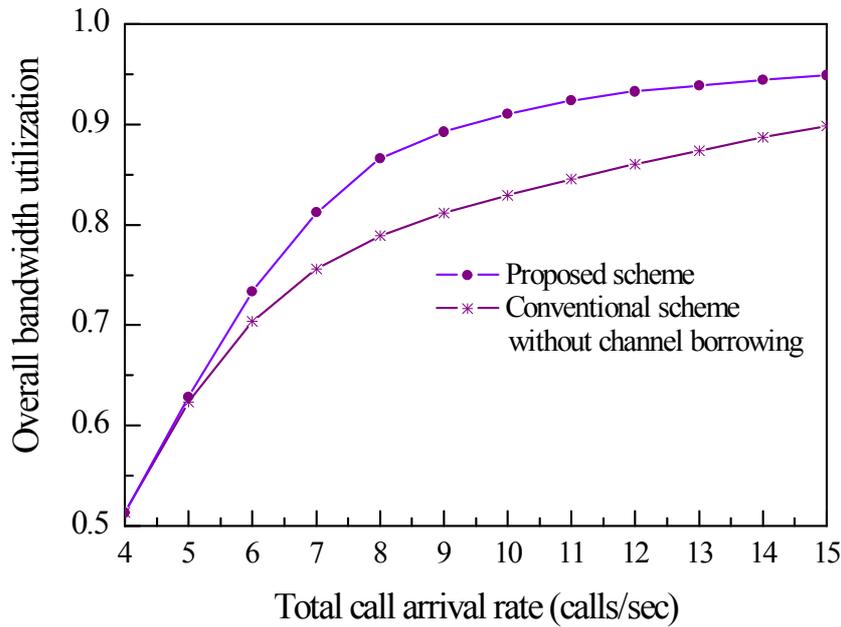

**Fig. 11.** Comparison of bandwidth utilization.

Maximum bandwidth utilization is mandatory for fruitful communication. The bandwidth utilization for the system based on the proposed scheme is maximized, shown in Fig. 11 that is compared to the conventional scheme i.e. scheme without channel borrowing process. The lower value of total call arrival rate prescribes small variation in efficiency for both schemes as the traffic intensity in the cells of the cluster is not high. In this case bandwidth utilization is less as there remain unused channels in the cells. With the increment of the total call arrival rate, the proposed scheme shows better performance compared to the conventional model which is one of the salient features of the proposed scheme. As the call arrival rate in the reference cell is increasing gradually, the unused channels of the adjacent cells are borrowed by the reference cell to fulfill the demand of the excessive users. As a result, the unused channels of the adjacent cells are utilized. Thus the proposed model assurances better bandwidth utilization.

Figure 12 shows the SINR levels of the proposed scheme and the conventional scheme i.e. scheme without interference management after dynamic channel borrowing process. The result signifies that the SINR level decreases with the increment of the distance between the BS and the MS of the reference cell for both schemes. The interference management process increases the SINR level which is very significant for any distance with compared to the SINR level of the conventional scheme. As the reference cell and the adjacent cells are bifurcated and the interfering channels are provided to the inner part users, the users using same frequency band, receive strong signal from the BS of the reference cell and trifling interference from the BS of the adjacent cells. Furthermore, the interfering channels of the adjacent cells are also locked if they are unoccupied. Thus, it also reduces the interference.

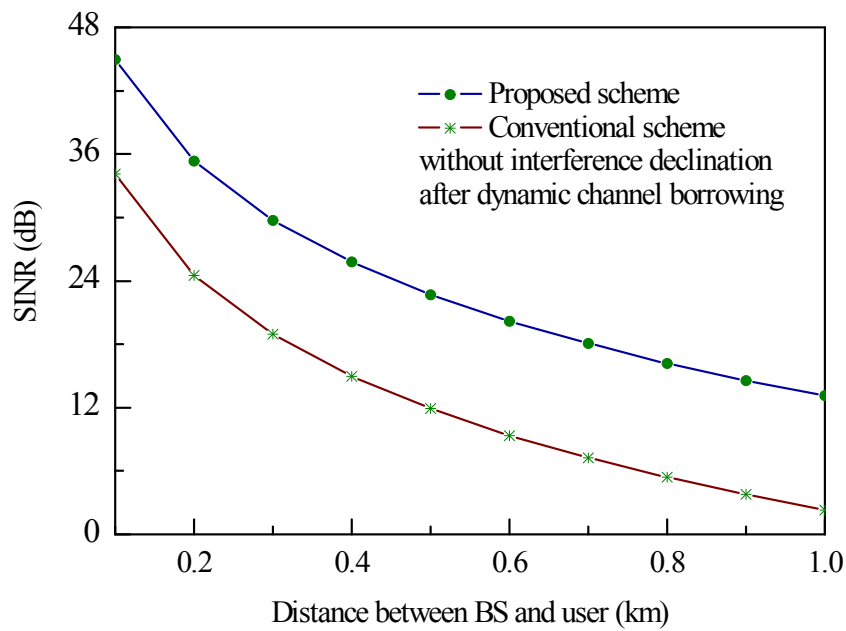

**Fig. 12.** Comparison of SINR levels in term of interference declination.

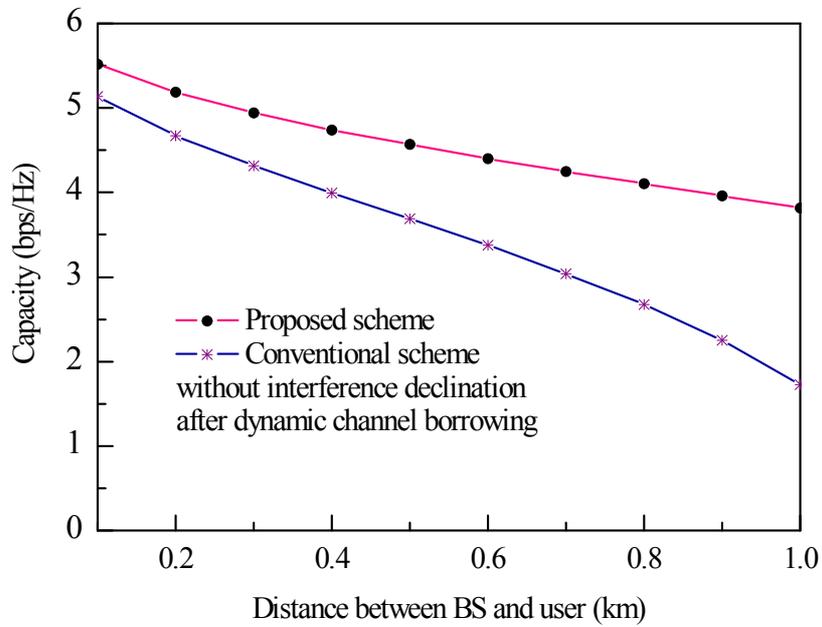

**Fig. 13.** Comparison of capacity in term of interference declination.

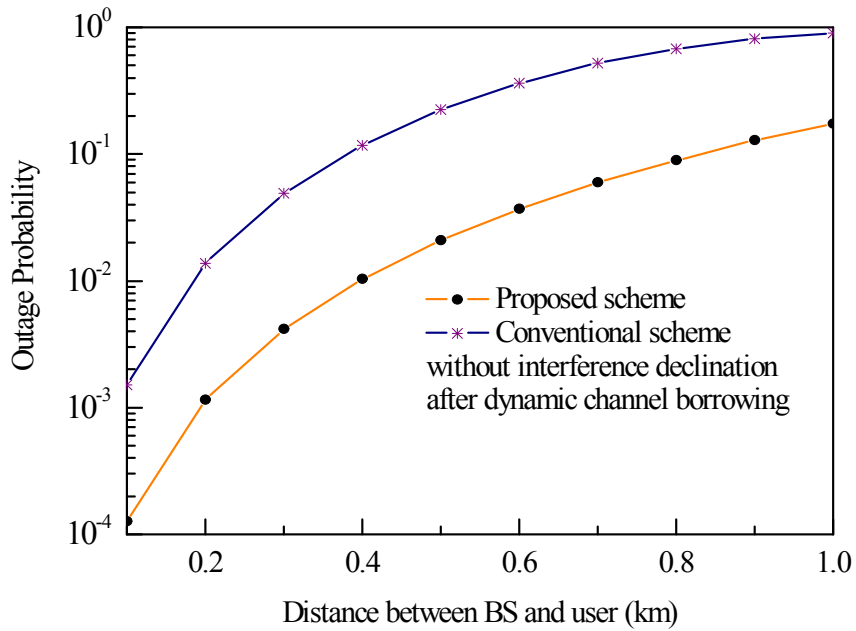

**Fig. 14.** Comparison of outage probability in case of interference declination.

The comparison of the capacity between the proposed scheme and conventional scheme i.e. scheme without interference management after dynamic channel borrowing process, is illustrated in Fig.13. For the analysis of the capacity, Shannon capacity formula is used. With the increment of the distance, the capacity of the network degrades slowly for the proposed model whereas the capacity diminishes rapidly in case of conventional scheme. As the SINR level of the proposed model is greater than that of conventional scheme, so the capacity of the network shows superior performance. Likewise, the proposed scheme shows a reliable data transmission rate than the conventional scheme without interference management.

Figure 14 illustrates that the outage probability of the proposed scheme is meaningfully smaller than the conventional method i.e. scheme without interference declination after dynamic channel borrowing. If the distance between BS and MS increases, the outage probability of the proposed scheme increases but the result remains in an acceptable range compared to the conventional scheme. The outage probability of the conventional scheme increases drastically as the users receive strong interference from adjacent cells when they move towards the edge of the cell because there is no interference management. The improved SINR level of the proposed scheme alleviates the outage probability of the network that proves smarter performance of the proposed model.

## 6. Conclusions

The main advantage of the proposed scheme is the use of same radio spectrum at the same time with insignificant interference and ensuring greater bandwidth utilization. To make it exclusively effective, different conditions are illustrated in this paper for interference declination so that it can draw the considerable interest for future wireless communication networks. In the previous time, there were lots of researches regarding interference management. Our proposed scheme shows three unique ways for interference management after the dynamic channel borrowing process. A deep scrutiny and simulation results prescribe the operational capability of the proposed scheme. The proposed scheme confirms reduced outage probability and overall call blocking probability without sacrificing bandwidth utilization. The scheme also provides outstanding performance in terms of system capacity and SINR level. Our proposed model is expected to be a better choice for the future wireless networks, considering the efficient performances. Our future research includes priority scheme, multiclass traffic, intercellular interference mitigation, inter-carrier interference management in OFDMA, interference declination in multicellular networks. We are also interested to implement the proposed model for femtocellular networks.

## References

[1] Mostafa Zaman Chowdhury, Yeong Min Jang, and Zygmunt J. Haas, "Call Admission Control Based on Adaptive Bandwidth Allocation for Wireless Networks," *IEEE/KICS Journal of Communications and Networks*, vol. 15, no. 1, pp. 15-24, February 2013.

[2] Mahmudur Rahman and Halim Yanikomeroglu, "Enhancing Cell-Edge Performance: A Downlink Dynamic Interference Avoidance Scheme with Inter-Cell Coordination," *IEEE Transactions on Wireless Communications*, vol. 9, no. 4, pp. 1414-1425, April 2010.

[3] Alireza Attar, Vikram Krishnamurthy, and Omid Namvar Gharehshiran, "Interference Management Using Cognitive Base-Stations for UMTS LTE," *IEEE Communications Magazine*, vol. 49, no. 8, pp. 152-159, August 2011.

[4] Mostafa Zaman Chowdhury, Yeong Min Jang, and Zygmunt J. Haas, "Radio Resource Allocation for Scalable Video Services over Wireless Cellular Networks," *Wireless Personal Communications*, vol. 74, no. 3, pp. 1061–1079, August 2013.

[5] Xian Wang, Pingzhi Fan, and Yi Pan, "A More Realistic Thinning Scheme for Call Admission Control in Multimedia Wireless Networks," *IEEE Transactions on Computers*, vol. 57, no. 8, pp. 1143-1148, August 2008.

[6] Salah Eddine Elayoubi, Olfa Ben Haddada, and Benoit Fouresti´, "Performance Evaluation of Frequency Planning Schemes in OFDMA-based Networks," *IEEE Transactions on Wireless Communications*, vol. 7, no. 5, pp. 1623-1633, May 2008.

[7] Syed Hussain Ali and Victor C. M. Leung, "Dynamic Frequency Allocation in Fractional Frequency Reused OFDMA Networks," *IEEE Transactions on Wireless Communications*, vol. 8, no. 8, pp. 4286-4295, August 2009.

[8] Zuoying Xu, Pitu B. Mirchandani, and Susan H. Xu, "Virtually Fixed Channel Assignment in Cellular Mobile Networks with Recall and Handoffs," *Telecommunication Systems*, vol. 13, no. 2-4, pp. 413-439, July 2000.

[9] Geetali Vidyarthi, Alioune Ngom, and Ivan Stojmenovic´, "A Hybrid Channel Assignment Approach Using an Efficient Evolutionary Strategy in Wireless Mobile Networks," *IEEE Transactions on Vehicular Technology*, vol. 54, no. 5, pp. 1887-1895, September 2005.

[10] Hua Jiang and Stephen S. Rappaport, "Channel Borrowing without Locking for Asynchronous Hybrid FDMA/TDMA Cellular Communications," *Wireless Personal Communication*, vol. 9, no. 3, pp. 233-254, May 1999.

[11] Kyuho Son and Song Chong, "Dynamic Association for Load Balancing and Interference Avoidance in Multi-Cell Networks," *IEEE Transactions on Wireless Communications*, vol. 8, no. 7, pp. 3566-3576, July 2009.

[12] Xunyong Zhang, Chen He, Lingge Jiang, and Jing Xu, "Inter-cell Interference Coordination Based on Softer Frequency Reuse in OFDMA Cellular Systems," In Proceeding of *IEEE International Conference Neural Networks & Signal Processing*, June 8, 2008, pp. 270-275.

[13] Xia Wang and Shihua Zhu, "Mitigation of Intercarrier Interference Based on General Precoder Design in OFDM Systems," In Proceeding of *International Conference on Advanced Information Networking and Applications*, May 27, 2009, pp. 705-710.

[14] Aravind Iyer, Catherine Rosenberg, and Aditya Karnik, "What is the Right Model for Wireless Channel Interference?," *IEEE Transactions on Wireless Communications*, vol. 8, no. 5, pp. 2662-2671, May 2009.

[15] Eunsung Oh, Seungyoup Han, Choongchae Woo, and Daesik Hong, "Call Admission Control Strategy for System Throughput Maximization Considering both Call and Packet-Level QoSs," *IEEE Transactions on Communications*, vol. 56, no. 10, pp. 1591-1595, October 2008.

[16] Hyun-Seung Kim, Deok-Rae Kim, Se-Hoon Yang, Yong-Hwan Son, and Sang-Kook Han, "Mitigation of Inter-Cell Interference Utilizing Carrier Allocation in Visible Light Communication System," *IEEE Communications Letters*, vol. 16, no. 4, pp. 526-529, April 2012.

[17] Mostafa Zaman Chowdhury, Seung Que Lee, Byung Han Ru, Namhoon Park, and Yeong Min Jang, "Service Quality Improvement of Mobile Users in Vehicular Environment by Mobile Femtocell Network Deployment," In Proceeding of IEEE International Conference on ICT Convergence, September 2011, pp. 194-198.

[18] Mostafa Zaman Chowdhury, Yeong Min Jang, and Zygmunt J. Haas, "Cost-Effective Frequency Planning for Capacity Enhancement of Femtocellular Networks," *Wireless Personal Communications*, vol. 60, no. 1, pp. 83–104, September 2011.

[19] Shaoji Ni, Yong Liang, and Sven-Gustav Häggman, "Outage Probability in GSM-GPRS Cellular Systems with and without Frequency Hopping," *Wireless Personal Communication*, vol. 14, no. 3, pp. 215-234, September 2000.

[20] Michael Gastpar, "On Capacity under Receive and Spatial Spectrum-Sharing Constraints," *IEEE Transaction Information Theory*, vol. 53, no. 2, pp. 471–487, February 2007.

# Biographies

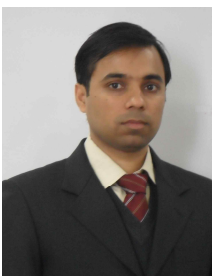

**Mostafa Zaman Chowdhury** received his B.Sc. degree in electrical and electronic engineering from Khulna University of Engineering and Technology (KUET), Bangladesh, in 2002. He received his M.Sc. and Ph.D. degrees both in electronics engineering from Kookmin University, Korea, in 2008 and 2012, respectively. In 2003, he joined the Electrical and Electronic Engineering Department at KUET as a faculty member. Currently he is working as an Assistant Professor at the same department. In 2008, he received the Excellent Student Award from Kookmin University. One of his papers received the Best Paper Award at the International Conference on Information Science and Technology in April 2012 in

Shanghai, China. He served as a reviewer for several international journals (including IEEE Communications Magazine, IEEE Transaction on Vehicular Technology, IEEE Communications Letters, IEEE Journal on Selected Areas in Communications, Wireless Personal Communications (Springer), Wireless Networks (Springer), Mobile Networks and Applications (Springer), and Recent Patents on Computer Science) and IEEE conferences. He has been involved in several Korean government projects. His research interests include convergence networks, QoS provisioning, mobility management, femtocell networks, and VLC networks.

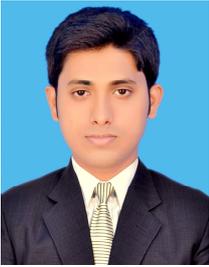

**Mohammad Arif Hossain** moved to Seoul, Korea from Bangladesh with a view to achieving his M.Sc. degree. He is pursuing his M.Sc. program in Electronics Engineering Department of Kookmin University, Korea and received his B.Sc. degree in Electrical and Electronic Engineering from Khulna University of Engineering and Technology (KUET), Bangladesh, in 2014. At this time he is working as a research assistant at the Wireless Network Lab (WNL) of Kookmin University, Korea and involving in Korean government projects. His research interests include visible light communication (VLC), optical communications for cameras (OCC), beacon networks, positioning in wireless networks, MIMO etc.

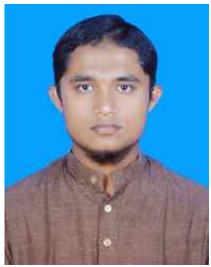

**Shakil Ahmed** received his B.Sc. degree in electrical and electronic Engineering from Khulna University of Engineering and Technology (KUET), Bangladesh, in 2014. Currently, he is researching on wireless networks. His research interests include radio resource allocation, femtocell networks, convergence network, and smart grid.

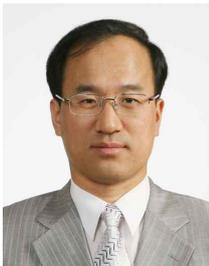

**Yeong Min Jang** received the B.E. and M.E. degrees both in electronics engineering from Kyungpook National University, Korea, in 1985 and 1987, respectively. He received the doctoral degree in Computer Science from the University of Massachusetts, USA, in 1999. He worked for ETRI between 1987 and 2000. Since September 2002, he is with the School of Electrical Engineering, Kookmin University, Seoul, Korea. He has organized several conferences such as ICUFN2009, ICUFN2010, ICUFN2011, ICUFN2012, and ICUFN2013. He is currently a member of the IEEE and a life member of KICS (Korean Institute of Communications and Information Sciences). He had been the director of the Ubiquitous IT Convergence Research Center at Kookmin University since 2005 and the director of LED Convergence Research Center at Kookmin University since 2010. He has served as the executive director of KICS since 2006. He had been the organizing chair of Multi Screen Service Forum of Korea since 2011. He had been the Chair of IEEE 802.15 LED Interest Group (IG-LED). He received the Young Science Award from the Korean Government (2003 to 2005). He had served as the founding chair of the KICS Technical Committee on Communication Networks in 2007 and 2008. His research interests include 5G mobile communications, radio resource management, small cell networks.